\documentclass{appolb}
\usepackage{epsfig}

\begin{document}

\title{STRUCTURE FORMATION\\ IN THE QUINTESSENTIAL UNIVERSE
\thanks{Presented at the XXV International School of Theoretical Physics 
``Particles and Astrophysics -- Standard Models and Beyond", Ustro\'n, Poland,
September 10-16, 2001.}
\thanks{Work supported in part by the Polish State Committee for Scientific
Research (KBN) grant No. 2P03D02319.}
}

\author{Ewa L. {\L}okas
\address{Nicolaus Copernicus Astronomical
Center\\ Bartycka 18, 00--716 Warsaw, Poland}}

\maketitle
\begin{abstract}
I review the main characteristics of structure formation in the quintessential
Universe. Assuming equation of state $w=p/\varrho=$const I provide a brief
description of the background cosmology and discuss the linear growth
of density perturbations, the  strongly nonlinear evolution, the power spectra
and rms fluctuations as well as mass functions focusing on the three values 
$w=-1, -2/3$ and $-1/3$. Finally I describe the presently available and future 
constraints on $w$.

\end{abstract}
\PACS{95.35.+d, 98.62.-g, 98.62.Ai, 98.62.Ck, 98.80.-k, 98.80.Es}

\section{Introduction}

Our knowledge of background cosmology has recently improved
dramatically due to new supernovae and cosmic microwave background data.
Current observations favor a flat Universe with matter density
$\Omega_0=0.3$ \cite{hr} and the remaining contribution in the
form of cosmological constant \cite{cpt, llpr}
or some other form of dark energy. The models with cosmological constant
are known, however, to suffer from two major problems. One is related to
the origin of the constant - it cannot be explained in terms of the vacuum
energy since its energy is orders of magnitude smaller. The other is the lack
of explanation why the present densities in matter and cosmological constant are
comparable.

A new class of models that solve these problems and also satisfy present
observational constraints has been proposed a few years ago \cite{cds}.
In these models the cosmological constant is replaced with a new
energy component, called quintessence, characterized by the equation of state
$p/\varrho=w \neq -1$. The component can cluster on largest scales
and therefore affect the mass power spectrum \cite{ma}
and microwave background anisotropies \cite{balbi, bacci}.

The investigations of the physical basis for the
existence of such component are now more than a decade old \cite{rp}.
One of the promising models is based on so-called
``tracker fields" that display an attractor-like behavior causing the
energy density of quintessence to follow the radiation density in the
radiation dominated era but dominate over matter density after
matter-radiation equality \cite{zws, swz}. It is still debated, however,
how $w$ should depend on time, and whether its redshift dependence can be
reliably determined observationally \cite{bm, mbs, wa}.

A considerable effort has gone into attempts to put constraints on models
with quintessence and presently the values of $-1 <w<-0.6$ seem most
feasible observationally \cite{wang, ht}. Here I review the main characteristics
of structure formation in the quintessential Universe which may provide constraints
on the equation of state. In the last Section I discuss the current status of
observational limits on $w$ and future perspectives.

\section{Background cosmology}

Quintessence obeys the following equation of state relating its density
$\varrho_Q$ and pressure $p_Q$
\begin{equation}    \label{q1}
    p_Q = w \varrho_Q, \ \ \ \ {\rm where} \ \ -1 \le w < 0.
\end{equation}
The case of $w=-1$ corresponds to the usually defined cosmological
constant.

\begin{figure}
\begin{center}
    \leavevmode
    \epsfxsize=12cm
    \epsfbox[50 50 590 310]{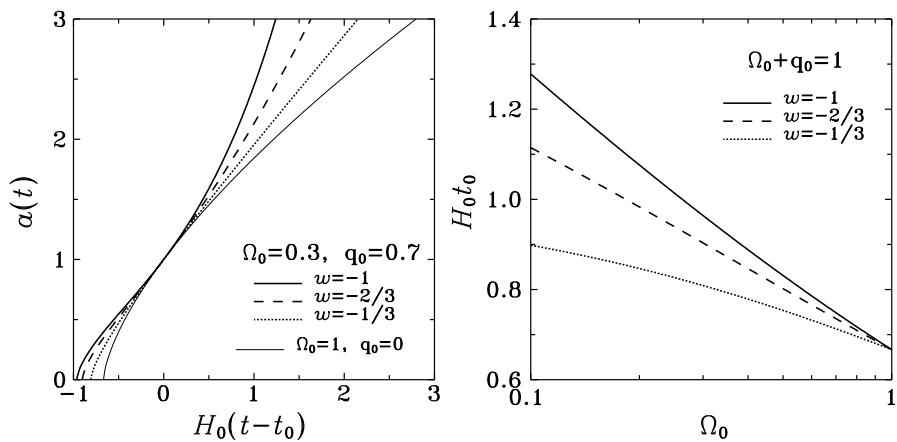}
\end{center}
    \caption{Left panel: Time evolution of the scale factor in different models.
    Right panel: The present age of Universe in units of $H_0^{-1}$
    in flat models as a function of $\Omega_0$ for different $w$.}
\label{aodtage}
\end{figure}

The evolution of the scale factor $a=R/R_0=1/(1+z)$ (normalized to unity
at present, $z$ is the redshift) in the quintessential Universe is governed by the
Friedmann equation
\begin{equation}       \label{th1}
    \frac{{\rm d} a}{{\rm d} t} = \frac{H_0}{u(a)}
\end{equation}
where
\begin{equation}    \label{th2}
    u(a) = \left[ 1+ \Omega_0 \left(\frac{1}{a}
    -1\right) + q_0 \left(\frac{1}{a^{1+3 w}} - 1\right) \right]^{-1/2}
\end{equation}
and $H_0$ is the present value of the Hubble parameter.
The quantities with subscript $0$ here and below denote the present
values. The parameter $\Omega$ is the standard measure of the amount of
matter in units of critical density and $q$
measures the density of quintessence in the same units
\begin{equation}    \label{q4}
    q  = \frac{\varrho_Q}{\varrho_{\rm crit}}.
\end{equation}
The Einstein equation for acceleration ${\rm d}^2 a/{\rm d} t^2 =
-4 \pi G a (p +\varrho/3)$ shows that $w < -1/3$ is needed for the
accelerated expansion to occur. The left panel of Fig.~\ref{aodtage} shows
the evolution of scale factor in different models. The right panel presents
the dependence on $w$ of the present age of Universe
\begin{equation}    \label{q5}
    t_0 = \frac{1}{H_0} \int_0^1 u(a) {\rm d} a.
\end{equation}

Solving the equation for the conservation of energy ${\rm d}
(\varrho_Q a^3)/{\rm d}a = -3 p_Q a^2$ with condition (\ref{q1}) we get the
evolution of the density of quintessence which for $w={\rm const}$, the case
considered in this paper, reduces to
\begin{equation}    \label{q3}
    \varrho_Q = \varrho_{Q,0} \ a^{-3(1+w)}.
\end{equation}

The evolution of $\Omega$ and $q$ with scale factor is given by
\begin{equation}   \label{th3}
    \Omega(a) = \frac{\Omega_0 u^2(a)}{a}, \ \ \ \ \ q(a) = \frac{q_0 u^2(a)}{a^{1+3 w}}
\end{equation}
while the Hubble parameter itself evolves so that $H(a) = H_0/[a \ u(a)]$.

\section{Linear growth of perturbations}

\begin{figure}
\begin{center}
    \leavevmode
    \epsfxsize=8cm
    \epsfbox[40 40 330 300]{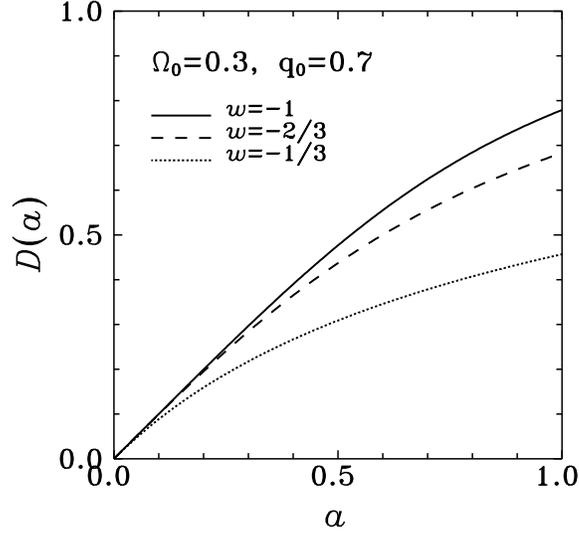}
\end{center}
    \caption{The linear growth rate of density fluctuations for
    $\Omega_0=0.3$, $q_0=0.7$ in three cases of $w=-1, -2/3$ and $-1/3$. }
\label{doda}
\end{figure}

The linear evolution of the matter density contrast $\delta=\delta
\varrho/\varrho$ is governed by equation \cite{pee}
\begin{equation}    \label{th5a}
    \ddot{\delta} + 2 \frac{\dot{a}}{a} \dot{\delta} - 4 \pi G \varrho
    \delta =0
\end{equation}
where dots represent derivatives with respect to time.
For flat models and arbitrary $w$ an analytical expression for $D(a)$,
the growing mode of the time-dependent part of $\delta$, was found
\cite{sw}. With our notation and the normalization of $D(a)=a$ for $\Omega=1$ and
$q=0$ it becomes
\begin{equation}    \label{th8}
    D(a) = a \ \ _2 F_1 \left[ -\frac{1}{3 w},
    \frac{w-1}{2 w}, 1-\frac{5}{6 w}, - a^{-3 w}
    \frac{1-\Omega_0}{\Omega_0} \right]
\end{equation}
where $_2 F_1 $ is a hypergeometric function.
The solutions (\ref{th8}) for different $w$ and cosmological parameters
$\Omega_0=0.3$ and $q_0=0.7$ are plotted in Fig.~\ref{doda}.

\begin{figure}
\begin{center}
    \leavevmode
    \epsfxsize=12cm
    \epsfbox[50 50 590 310]{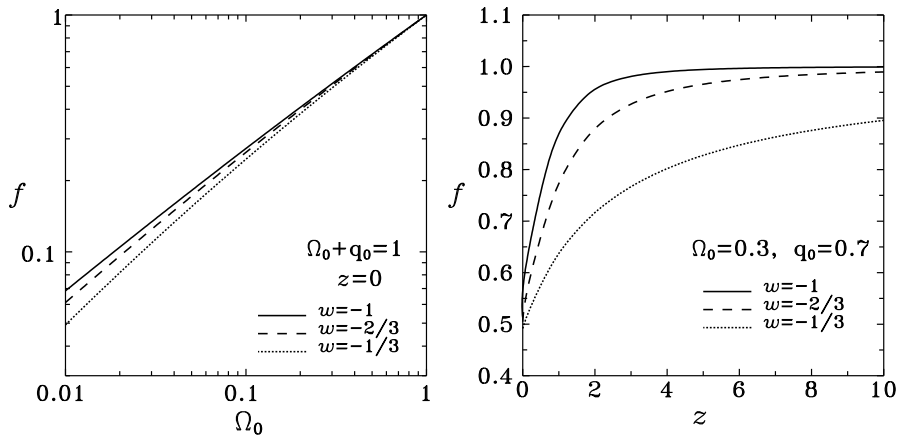}
\end{center}
    \caption{Left panel: The velocity factor $f$ at present as a function of
    $\Omega_0$ for flat models with different $w$. Right panel: The redshift
    dependence of $f$ for $\Omega_0=0.3$, $q_0=0.7$ and different $w$.}
\label{ffz}
\end{figure}

The peculiar velocity field in linear perturbation theory is obtained from \cite{pee}
\begin{equation}    \label{th8a}
    {\bf v} = \frac{2 f}{3 \Omega H} {\bf g}
\end{equation}
where ${\bf g}=- \nabla \phi/a$ is the peculiar gravitational acceleration and $f$
is the dimensionless velocity factor
\begin{equation}    \label{th8b}
    f = \frac{a}{\dot{a}} \frac{\dot{D}}{D}.
\end{equation}
For flat models this formula can be evaluated analytically using Eq. (\ref{th8}).
The dependence of $f$ on $\Omega_0$ at present ($z=0$) for flat models with
different $w$ is shown in the left panel of Fig.~\ref{ffz}. We see immediately that
the dependence on $w$ is very weak. However, as shown in the right panel of
Fig.~\ref{ffz}, when going to higher redshifts we find
that the velocity factor is much more sensitive to $w$ which gives some hope for
applying it to determine $w$ from local peculiar velocity field.

\section{Strongly nonlinear evolution}

\begin{figure}
\begin{center}
    \leavevmode
    \epsfxsize=12cm
    \epsfbox[50 50 590 310]{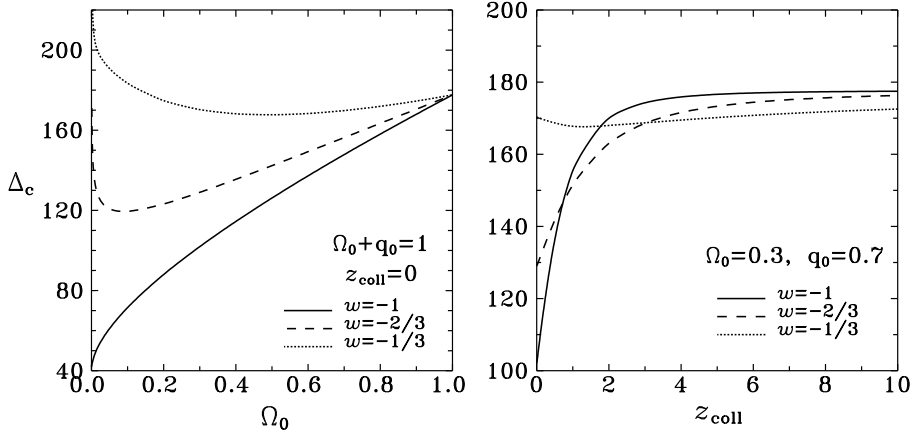}
\end{center}
    \caption{Left panel: dependence of $\Delta_{\rm c}$ on
    $\Omega_0$ and $w$ for flat models and collapse time
    $z_{\rm coll}=0$. Right panel: $\Delta_{\rm c}$ as a function of $z_{\rm coll}$ for
    $\Omega_0=0.3$, $q_0=0.7$ and different $w$.}
\label{dlwwz}
\end{figure}

The simplest model of formation of bound objects (called the
spherical or top hat model) \cite{gg, g, lha, lhb}
describes the nonlinear evolution of a uniform spherical density
perturbation in an otherwise homogeneous Universe. If the overdensity is
big enough the perturbed region will slow down the expansion and eventually turn around
and collapse. The evolution of the size of such a perturbation can be described 
\cite{ws} by the Einstein equation
\begin{equation}	\label{w1}
    \frac{ \ddot{r}}{r} = - 4 \pi G \left[\left(w + \frac{1}{3} \right) \varrho_Q 
    + \frac{1}{3} \varrho \right].
\end{equation}
Together with the Friedmann equation (\ref{th1}) it can be solved to obtain the 
ratio of the cluster density to the background density at the time of turn-around
\begin{equation}	\label{w2}
    \zeta =\frac{\varrho}{\varrho_{\rm b}}(z_{\rm ta}) = 
    \left(\frac{3 \pi}{4} \right)^2 \Omega^{-0.79 + 0.26 \Omega -0.06 w} |_{z_{\rm ta}},
\end{equation}
where the last formula is the approximation given in \cite{ws}. 

One of the most useful quantities one can derive from the spherical top hat model is
the density of the virialized halo in units of the critical density at 
the time of collapse corresponding to redshift $z_{\rm coll}$
\begin{equation}	\label{w3}
    \Delta_{\rm c} = \frac{\varrho}{\varrho_{\rm crit}} (z_{\rm coll}) = \zeta(z_{\rm ta})
    \Omega (z_{\rm coll}) \left(\frac{r_{\rm ta}}{r_{\rm coll}} \right)^3 
    \left( \frac{1+z_{\rm ta}}{1+z_{\rm coll}} \right)^3.
\end{equation}
The redshifts $z_{\rm coll}$ and $z_{\rm ta}$ are related by the assumption that 
the collapse time $t_{\rm coll}$, the time corresponding to $r \rightarrow
0$, is twice the turn-around time
\begin{equation}    \label{w4}
    t_{\rm ta} = \frac{1}{H_0}   \int_0^{a_{\rm ta}} u(a) {\rm d} a.
\end{equation}
In reality the object does not decrease its size to zero but virializes to reach the
effective final radius $r_{\rm coll}$.
The ratio by which the halo collapses is estimated from the virial theorem which 
gives \cite{ws}
\begin{equation}    \label{w5}
    \frac{r_{\rm coll}}{r_{\rm ta}} = \frac{1-\eta_v/2}{2+\eta_t - 3 \eta_v/2},
\end{equation}
where
\begin{eqnarray}    
    \eta_t &=& \frac{2}{\zeta} \frac{q(z_{\rm ta})}{\Omega(z_{\rm ta})}  \label{w6} \\
    \eta_v &=& \frac{2}{\zeta} \frac{q(z_{\rm coll})}{\Omega(z_{\rm coll})}
    \left( \frac{1+z_{\rm coll}}{1+z_{\rm ta}} \right)^3.  \label{w7}
\end{eqnarray}
Figure~\ref{dlwwz} shows how $\Delta_{\rm c}$ depends on $\Omega_0$, $w$ and the redshift 
of collapse.

\section{Power spectrum of density fluctuations}

Power spectrum $P(k, a)$ is defined as the Fourier transform of the correlation
function of density fluctuations
\begin{equation}   \label{q14a}
    P(k,a) = \int \xi(r,a) \ {\rm e}^{-{\rm i}{\bf k} \cdot {\bf r}} {\rm d}^3 r.
\end{equation}
The spectra for Universe dominated by cold
dark matter (CDM) have been widely discussed in the literature, e.g. \cite{sugi}.
For the present time $(a=1)$ the power spectrum is usually written in the form
\begin{equation}   \label{q15}
    P(k) = A k^n T^2 (k)
\end{equation}
where $n$ measures the slope of the primordial power spectrum (we will
assume $n=1$), $T$ is the transfer function and $A$ is a normalization constant.
In the presence of cosmological constant ($\Lambda$CDM) the transfer function
$T_\Lambda$ can be approximated by \cite{sugi}
\begin{equation}   \label{q15a}
   T^2_{\Lambda}(p) = \frac{\ln^2(1+2.34 p)}{(2.34 p)^2}[1 + 3.89 p + (16.1 p)^2 
   + (5.46 p)^3 + (6.71 p)^4]^{-1/2},
\end{equation}
where $p=k/(\Gamma h {\rm Mpc}^{-1})$, $\Gamma=\Omega_0 h \exp [-\Omega_{\rm b}
(1+ \sqrt{2 h}/\Omega_0)]$ and $h=H_0/[100$ km/(s Mpc)] $=0.7$. We assume the
barion contribution to the density parameter $\Omega_{\rm b}=0.04$.

\begin{figure}
\begin{center}
    \leavevmode
    \epsfxsize=12cm
    \epsfbox[50 50 590 310]{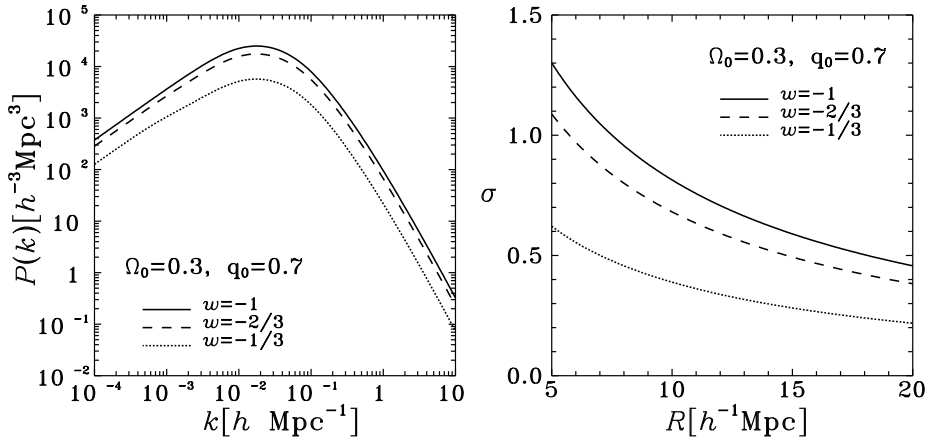}
\end{center}
    \caption{The linear power spectra (left panel) and rms fluctuation as a function
    of smoothing scale (right panel) for flat models with different $w$.}
\label{pofksig}
\end{figure}

For flat models with quintessence the modification of
the transfer function has been proposed by Ma {\em et al.} \cite{ma}. The transfer
function in equation (\ref{q15}) is then $T_Q = T_{Q\Lambda} T_\Lambda$, where
$T_{Q\Lambda} = T_Q/T_\Lambda$ is approximated by fits given in \cite{ma}.
The present linear power spectra obtained for $w=-1, -2/3$ and $-1/3$ assuming 
COBE normalization are shown in the left panel of Fig.~\ref{pofksig}.

The rms density fluctuation, $\sigma$, at comoving smoothing scale $R$ is given by
\begin{equation}   \label{q13}
    \sigma^2 = \frac{1}{(2 \pi)^3} \int {\rm d}^3 k P(k)
    W^2_{TH}(k R)
\end{equation}
where the smoothing is performed with the top hat filter
\begin{equation}   \label{q14}
    W^2_{TH}(k R) = \frac{3}{(k R)^2} \left( \frac{\sin k R}{k R} - \cos
    k R \right).
\end{equation}
The dependence of $\sigma$ on smoothing scale for flat models with different $w$
is shown in the right panel of Fig.~\ref{pofksig}. A particularly useful quantity,
constrained by cluster abundance is the rms fluctuation at the scale of $8 h^{-
1}$ Mpc. Due to COBE normalization its values turn out to depend strongly on $w$ and we get
$\sigma_8 =0.96, 0.80$ and $0.46$ for $w=-1, -2/3$ and $-1/3$ respectively.

\section{Mass functions}

One of the most important measures of structure formation is provided by the mass
function of bound objects. Using the analytical prescription of Press and Schechter
\cite{ps}, we can estimate the
cumulative mass function (the comoving number density of objects of mass
grater than $M$)
\begin{equation}   \label{q12a}
    N(>M) = \int_M^\infty n(M) {\rm d} M
\end{equation}
where $n(M)$ is the number density of objects with mass between $M$ and
$M+{\rm d}M$
\begin{equation}   \label{q12}
    n(M) = - \left( \frac{2}{\pi} \right)^{1/2} \frac{\varrho_{\rm b}}{M}
    \frac{\delta_{\rm c}}{\sigma^2} \frac{{\rm d} \sigma}{{\rm d} M}
    \exp \left( - \frac{\delta_{\rm c}^2}{2 \sigma^2} \right).
\end{equation}
In the expression above, $\varrho_{\rm b}$ is the background density, $\sigma$
is the rms density fluctuation at comoving smoothing scale $R$ described
in Section~5. The mass is related to the smoothing scale by $M=4 \pi \varrho_{\rm
b} R^3/3$. The parameter $\delta_{\rm c}$ is the halo density
at collapse as predicted by linear theory. We adopt here its standard value 
$\delta_{\rm c}=1.69$ (exact for the Einstein-de Sitter Universe) since it depends 
very weakly on $\Omega_0$ and $w$ \cite{ws}.

\begin{figure}
\begin{center}
    \leavevmode
    \epsfxsize=8cm
    \epsfbox[40 40 330 300]{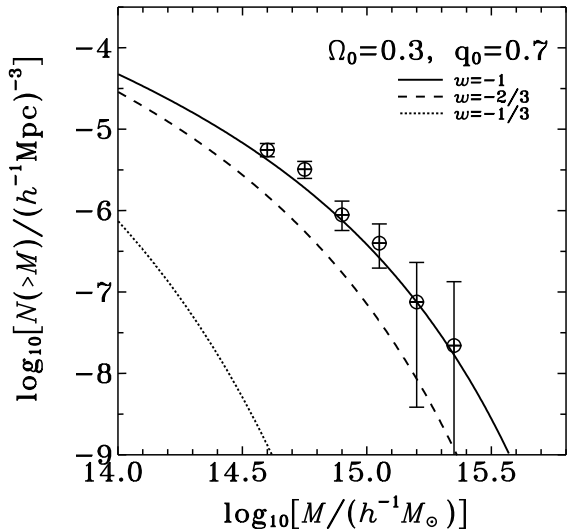}
\end{center}
    \caption{The Press-Schechter cumulative mass functions for different $w$
    assuming $\Omega_0 = 0.3$ and $q_0=0.7$.}
\label{cmfth}
\end{figure}

Figure~\ref{cmfth} shows the cumulative mass functions calculated from equation
(\ref{q12}) with $\Omega_0 = 0.3$ and $q_0=0.7$ for three models with $w=-
1, -2/3$ and $-1/3$. The strong dependence on $w$ is due to the COBE normalization
of the power spectra. The Figure also shows data for rich clusters of galaxies from
\cite{girardi}. For our choice of
cosmological parameters the value of $w \approx -1$ is preferred.
However, the Press-Schechter formulae are known to underestimate the mass functions
on the scale of clusters of galaxies when compared to N-body simulations. The mass 
functions are also very sensitive to the value of $\Omega_0$. Therefore
when more exact predictions are used and $\Omega_0$ is measured to be close to $0.3$ with
a small error bar, we can expect higher values of $w$ to be preferred by the cluster data.

\section{Constraints on $w$}

The primary constraint on the value of $w$ comes from the observations of accelerating
expansion, which can only be obtained in models with $w<-1/3$. Age of Universe is
quite sensitive to $w$ and increases for lower $w$, however the accuracy of
our knowledge of both $H_0$ and $t_0$
is not good enough to put strong constraints on $w$. Current estimates are consistent
with $w<-0.5$.

One of the strongest arguments for the existence of dark energy comes from the studies
of cosmic microwave background (CMB). Although its power spectrum is weakly sensitive to
$w$ (e.g. the height of the first acoustic peak is somewhat increased and its position is
shifted to higher multipoles for lower $w$), it has already provided some limits.
Combining the data from COBE and recent balloon experiments Balbi {\em et al.}
\cite{balbi} find $-1<w<-0.6$ while Baccigalupi {\em et al.} \cite{bacci}
estimate the best-fitting value of $w$ to be $w=-0.8$. The ongoing
and future satellite experiments are expected to put even stronger limits on $w$.

Among the promising probes of dark energy are also supernovae Ia. The comoving
distance they serve to measure is sensitive only to the interesting cosmological
parameters and the errors related to supernova evolution or extinction are
estimated to be small. The existing data restrict $w$ only weakly \cite{ptw}, but future
experiments like The Supernova Acceleration Probe are expected to measure
the value of $w$ with a few percent accuracy \cite{nw}.

Structure formation also offers methods to constrain the cosmic equation of state.
The suppression of linear growth of density fluctuations for higher $w$ alone shows
that only for $w<-1/2$ the structure observed today could have evolved from small initial
perturbations deduced from CMB observations. The same range of acceptable $w$ values
follows from the behaviour of $\sigma_8$ which is a strongly decreasing
function of $w$.

The most promising tests are based on the number counts of galaxy
clusters. It turns out \cite{ws} that the slope of comoving abundance as a
function of redshift depends sensitively on $w$ and therefore can be used
to break degeneracies between $w$
and other cosmological parameters that appear e.g. in the analysis of CMB.
Such measurements are expected to be performed using the proposed new X-ray and
Sunyaev-Zeldovich effect surveys \cite{hmh} and the ongoing DEEP Redshift
Survey \cite{nd}. The constraints from structure formation appear to be
complementary to those from supernovae and CMB measurements.


\begin{thebibliography}{99}


\bibitem{hr} S. M. Harun-or-Rashid, M. Roos, {\em Astron. \& Astrophys.},
    {\bf 373}, 369 (2001)
\bibitem{cpt} S. M. Carroll, W. H. Press, E. L. Turner,
    {\em Annual Review of Astron. \& Astrophys.}, {\bf 30}, 499 (1992)
\bibitem{llpr} O. Lahav, P. B. Lilje, J. R. Primack, M. J. Rees,
    {\em MNRAS}, {\bf 251}, 128 (1991)
\bibitem{cds} R. R. Caldwell, R. Dave, P. J. Steinhardt,
   Phys. Rev. Lett., {\bf 80}, 1582 (1998)
\bibitem{ma} C. P. Ma, R. R. Caldwell, P. Bode, L. Wang,
    {\em Astrophys. J.}, {\bf 521}, L1 (1999)
\bibitem{balbi} A. Balbi, C. Baccigalupi, S. Matarrese, F. Perrotta, N. Vittorio,
    {\em Astrophys. J. Lett.}, {\bf 547}, 89 (2001)
\bibitem{bacci} C. Baccigalupi, A. Balbi, S. Matarrese, F. Perrotta, N. Vittorio,
   {\em Phys. Rev. D}, {\bf 65}, 63520 (2002)
\bibitem{rp} B. Ratra, P. J. E. Peebles, {\em Phys. Rev. D}, {\bf 37}, 3406 (1988)
\bibitem{zws} I. Zlatev, L. Wang, P. J. Steinhardt, {\em Phys. Rev.
    Lett.}, {\bf 82}, 896 (1999)
\bibitem{swz} P. J. Steinhardt, L. Wang, I. Zlatev, {\em Phys. Rev. D},
    {\bf 591}, 270 (1999)
\bibitem{bm} V. Barger, D. Marfatia, {\em Phys. Lett. B}, {\bf 498}, 67 (2001)
\bibitem{mbs} I. Maor, R. Brustein, P. J. Steinhardt, {\em Phys. Rev.
    Lett.}, {\bf 86}, 6 (2001)
\bibitem{wa} J. Weller, A. Albrecht, {\em Phys., Rev. Lett.}, {\bf 86}, 1939 (2001)
\bibitem{wang} L. Wang, R. R. Caldwell, J. P. Ostriker,P. J. Steinhardt,
    {\em Astrophys. J.}, {\bf 530}, 17 (2000)
\bibitem{ht} D. Huterer, M. S. Turner, {\em Phys. Rev. D}, {\bf 64}, 123527 (2001)
\bibitem{pee} P. J. E. Peebles,  {\em Principles of Physical Cosmology},
    Princeton Univ. Press, Princeton 1993
\bibitem{sw} V. Silveira, I. Waga, {\em Phys. Rev. D}, {\bf 50}, 4890 (1994)
\bibitem{gg} J. E. Gunn, J. R. Gott, {\em Astrophys. J.}, {\bf 176}, 1 (1972)
\bibitem{g} J. E. Gunn, {\em Astrophys. J.}, {\bf 218}, 592 (1977)
\bibitem{lha} E. L. \L okas, Y. Hoffman, {\em The Identification of
    Dark Matter}, Proc. 3rd International Workshop, eds N. J. C. Spooner,
    V. Kudryavtsev, World Scientific, Singapore 2001, p. 121
\bibitem{lhb} E. L. \L okas, Y. Hoffman, astro-ph/0108283
\bibitem{ws} L. Wang, P. J. Steinhardt, {\em Astrophys. J.}, {\bf 508}, 483 (1998)
\bibitem{sugi} N. Sugiyama, {\em Astrophys. J.}, {\bf 471}, 542 (1995)
\bibitem{ps} W. H. Press, P. Schechter, {\em Astrophys. J.}, {\bf 187}, 425 (1974)
\bibitem{girardi} M. Girardi, S. Borgani, G. Giuricin, F. Mardirossian,
	M. Mezzetti, {\em Astrophys. J.}, {\bf 506}, 45 (1998)
\bibitem{ptw} S. Perlmutter, M. S. Turner, M. White, {\em Phys., Rev. Lett.},
    {\bf 83}, 670 (1999)
\bibitem{nw} S. C. C. Ng, D. L. Wiltshire, {\em Phys. Rev. D}, {\bf 64}, 123519 (2001)
\bibitem{hmh} Z. Haiman, J. J. Mohr, G. P. Holder, {\em Astrophys. J.}, {\bf 553},
    545 (2001)
\bibitem{nd} J. A. Newman, M. Davis, {\em Astrophys. J. Lett.}, {\bf 534}, L11 (2000)

\end{thebibliography}
\end{document}